# The Black-Scholes-Merton dual equation

Shuxin GUO
*School of Economics and Management*
*Southwest Jiaotong University*
*Chengdu, Sichuan, P. R. China*
shuxinguo@home.swjtu.edu.cn
ORCID: 0000-0001-8188-6517

Qiang LIU*
*Institute of Chinese Financial Studies*
*Southwestern University of Finance and Economics*
*Chengdu, Sichuan, P. R. China.*
qiangliu@swufe.edu.cn
ORCID: 0000-0001-8466-3108

* corresponding author.

## Acknowledgments

We genuinely thank Steven Li for helping make the proof of Proposition 2a better. We are grateful to the 5th China Derivatives Youth Forum for the best paper award. This work was supported by the National Natural Science Foundation of China under Grant numbers 71701171, 72073109, and 72071162, by the Liberal Arts and Social Sciences Foundation of the Chinese Ministry of Education (21XJC790003), and by Huaxi Futures Co., Ltd.

**The Black-Scholes-Merton dual equation**

**Abstract**: We derive the Black-Scholes-Merton dual equation, which has exactly the same form as the Black-Scholes-Merton equation. The novel and general equation works for options with a payoff of homogeneous of degree one, including European, American, Bermudan, Asian, barrier, lookback, etc., leading to new insights into pricing and hedging. Perceptibly, a put-call equality emerges − all the put options can be priced as their corresponding calls by simultaneously swapping stock price (dividend yield) for strike price (risk-free rate), and vice versa. Equally important, we provide simple analytic formulas for hedging parameters delta and gamma, which elevate the put-call equality to true practical applicability. For futures options, the put-call equality leads to "symmetric" properties between puts and calls or among puts (calls).

## 1. INTRODUCTION

In his groundbreaking paper, Merton (1973) elegantly proves that call prices of American options are homogeneous of degree one.[1] The price homogeneity of derivatives turns out to be quite general for scale-invariant underlying processes, and the prices of many well-known options, such as Bermudan, Asian, and barrier, are actually homogeneous of degree one (Alexander and Nogueira, 2006).

Utilizing the price homogeneity and the Black-Scholes-Merton (BSM) differential equation (Black and Scholes, 1973; Merton, 1973),[2] we derive a second-order partial differential equation with respect to the option's strike as the fundamental variable. Amazingly, this new equation has exactly the same form as the celebrated BSM equation, but with the strike price replacing the underlying price and the risk-free interest rate switching places with the dividend yield. Because of the apparent "symmetry" between the new and BSM equations, we call it the Black-Scholes-Merton dual equation.[3] We believe, to the best of our knowledge, this novel equation has not been known before.

Importantly, the BSM dual equation is very general. Because the option's price preserves the homogeneity of its payoff (Alexander and Nogueira, 2006), the BSM dual equation works for any option with a payoff of homogeneous of degree one. Such options include, but are not limited to, European, American, Bermudan, Asian, barrier, and lookback. Note that the dual equation is utilized to make the Wu-Zhu static hedge of European options more accurate and extend the hedge to other options with a payoff of homogeneous of degree one (Guo and Liu, 2023).

[1] A function $f$ is homogeneous of degree one, if $f(ax, ay) = af(x, y)$ for all real $a > 0$.

[2] According to Alexander and Nogueira (2006), the geometric Brownian motion is scale-invariant.

[3] Due to the "homogeneity coupling" between the dual and BSM equations, Joseph Pimbley suggests calling it the BSM homogeneous-pair or H-pair equation (Guo and Liu, 2023).

The BSM dual equation has intriguing implications for the pricing of complex options. We prove perceptively that the price of put options is equal to that of call options via the dual equation, and vice versa. Equally important, the delta of put options turns out to be a simple analytic function of that of call options via the dual equation, as does gamma, and vice versa. As simple as it can be, the put-call price equality is accomplished by swapping stock price (dividend yield) for strike price (risk-free rate) simultaneously, which is readily verifiable in Section 3 for European options via the BSM formulas and for American options via binomial-tree pricing. This put-call equality is arguably more general than the classic put-call parity, which is valid solely for European options.

The BSM dual equation leads to further new insights on pricing, among which we show two. It is generally believed that American puts can only be priced numerically, unlike American calls. We show numerically that American currency puts can be approximated quite well by the Black-Scholes-Merton put formula, if the foreign risk-free rate is higher than the domestic risk-free rate. In the limit of a zero domestic risk-free rate, the BSM put formula becomes the exact price for American currency puts.

The second insight concerns the issue of known dollar dividends in pricing European options. Hull's popular textbook (2015) makes use of the risky component of the stock process. However, we argue that the risky component approach may be incorrect for American calls under the put-call equality. Therefore, the textbook approach to dollar dividends for European options pricing is probably problematic.

For futures options, the put-call equality results in novel relationships between put prices and call prices or among calls (puts). Because futures can be regarded as the underlying that pays a continuous dividend yield, which is exactly equal to the risk-free interest rate, the prices of puts and calls are "symmetric" in certain ways. We show three such symmetric properties. First, the price of put with strike $K$ at the underlying price $m^2S_0$ is $m$ times the price of call with strike $K$ at the underlying price $S_0$, where $m = \frac{K}{S_0}$. Second, the put and call have the same price, when the futures price and strike price switch roles. Third, the price of call with strike $K$ at the underlying price $m^2S_0$ is $m$ times the price of call with strike $S_0$ at the underlying price $K$. For all three cases, the deltas and gammas are also related via analytical equations, respectively.

We contribute to the derivatives' literature in three ways. First, the BSM dual equation and the put-call equality provide new perspectives on and innovative approaches to options' pricing, and are shown to shed light on several complex issues in options' pricing. Second, analytic formulas for delta and gamma elevate the put-call equality to true practical applicability and lead to simpler and more efficient implementations of options' pricing for real-world applications. Third, the BSM dual equation, which complements the BSM equation, is perhaps the last piece to complete the Black-Scholes-Merton framework.

Note that the BSM dual equation is not the only second-order partial differential equation with respect to the option's strike. Widely used as local volatility models, the Dupire equation (1993) and its generalized version (Gatheral, 2002) also utilize the strike price as the fundamental variable.[4] Important differences remain, however. Unlike our dual equation, Dupire is not scale-invariant (Alexander and Nogueira, 2006) and, in principle, works only for European options at maturity. Further, the generalized Dupire, which turns out to be the same as the dual equation (Guo and Liu, 2023), can be regarded as a special case of the dual equation.[5] Carr and Hirsa (2003) derive a partial integro differential equation with jumps for American options, which reduces to our dual equation if jumps are ignored, but is not scale-invariant in itself either and does not lead to the pricing insights that are the salient feature of our dual equation. Lastly, it is worth noting that the BSM dual equation is the most general.

---

[4] According to Breeden and Litzenberger (1978), the risk-neutral probability density at an option's maturity is a linear function of the second-order partial derivative of the price of a European call with respect to its strike. This relationship is utilized to derive both the Dupire equation and its generalized version.

[5] Being a special case of the dual equation, the (genelized) Dupire equation seems to have been interpreted incorrectly in the literature.

Also, relationships between puts and calls exist in the literature. Andreasen and Carr (2002) provide an excellent review of various ways of relating puts to calls, such as put-call equivalence (Grabbe, 1983; Schroder, 1999), put-call symmetry (Bates, 1988; Detemple, 1999), and put-call duality (Andreasen, 1997; Haug, 2002; Peskir and Shiryaev, 2002). Without a differential equation like the BSM dual equation, these papers work with the (expected) payoff of options or adopt advanced changes of numeraire; except Detemple (1999), all the others focus on European (and maybe American) options. Among them, the pricing results of the put-call symmetry (Detemple, 1999) for quite a few cases are identical to ours. Our novel approach stands out in four aspects, though. First, the dual equation provides a unifying framework for relating call and put options and includes most previous results as special cases. Second, the proof of the put-call equality instead utilizes the price property of a broad class of options directly via the dual equation, and thus is truly general, significantly simpler, and undoubtedly unique. Third, we derive simple analytic formulas for delta and gamma under the put-call equality, which are innovative and also unique. More importantly, the two greeks formulas make the BSM dual equation truly practical, because in derivatives' applications hedging (with delta and maybe gamma) and pricing can be regarded as the two sides of the same coin. Fourth, our dual reasoning leads to additional new insights into issues concerning options' pricing.

The remaining part of the paper is outlined here. Section 2 derives the Black-Scholes-Merton dual equation for options with payoffs of homogenous of degree one, and discusses its economic implications. Next in Section 3, we discuss the put-call equality, a new insight from the BSM dual equation, and several issues on options' pricing. Three symmetric price properties for futures options are discussed in Section 4. Finally, we conclude with remarks.

## 2. THE BSM DUAL EQUATION: A UNIFYING FRAMEWORK

Wu and Zhu (2016) proposed the ingenious scheme of hedging an option statically with a portfolio of options of the same kind but of different terms of maturity. The weights of the hedging options are determined by matching payoffs at the end of the hedging period via the approximate Dupire equation (1993). Researching on an improvement of the Wu-Zhu scheme, we serendipitously stumble onto the "symmetric" twin form of the celebrated Black-Scholes-Merton (BSM) equation.[6] To the best of our knowledge, the BSM twin form seems to have not been discussed in previous literature. This new twin form is presented in Proposition 1.

**PROPOSITION 1:** Under the Black-Scholes-Merton framework, the price of any option with a payoff of homogeneous of degree one satisfies the following Black-Scholes-Merton dual equation:

---

[6] Loosely, we use the term "BSM equation" to refer to the generalized equation with a continuous dividend yield or for a foreign currency (Garman and Kohlhagen, 1983).

$$\frac{\partial v}{\partial t} + (q - r)K\frac{\partial v}{\partial K} + \frac{1}{2}\sigma^2 K^2 \frac{\partial^2 v}{\partial K^2} = qv \tag{1}$$

where *v* is the option price, *q* is the continuous dividend yield, *r* is the risk-free interest rate, *K* is the strike price, and $\sigma$ is the volatility.

**PROOF**: According to Alexander and Nogueira (2006), the price of options preserves the homogeneity of their payoffs, given a scale-invariant underlying process; further, "geometric Brownian motions, even with mean reversion or a diffusion displacement, are" scale-invariant. Therefore, the prices of options with a payoff of homogeneous of degree one are homogeneous of degree one under BSM, too.

Denote the prices of such options by $v(S_t, K)$, where $S_t$ is the underlying price and *K* is the strike price. According to Euler's homogeneous function theorem, we have:

$$v = S_t \frac{\partial v}{\partial S_t} + K\frac{\partial v}{\partial K} \tag{2}$$

Furthermore, it is not difficult to show (see Appendix 1) that:

$$S_t^2 \frac{\partial^2 v}{\partial S_t^2} = K^2 \frac{\partial^2 v}{\partial K^2} \tag{3}$$

Combining Equations (2-3) with the Black-Scholes-Merton equation (with a constant dividend yield), we obtain Equation (1). QED.

One can readily see that Equation (1) has exactly the same form as the BSM differential equation with a continuous dividend yield (Black and Scholes, 1973; Merton, 1973), but with *K* substituting $S_t$ and *r* switching places with *q*. Note that the derivation is independent of *q*, *r*, or $\sigma$, so Equation (1) is not limited to these three parameters being constant.

Equation (1) is quite general. The dual equation works for a broad class of well-known options under BSM, such as European, American, Bermudan, barrier, and Asian, because their payoffs (and thus prices) are homogeneous of degree one (Alexander and Nogueira, 2006). It is worth mentioning that the dual equation is used to make the Wu-Zhu static hedge of European options more accurate and extend the hedge to options other than European ones (Guo and Liu, 2023). Note that the original BSM equation, which applies to all payoffs, is more general.

We arrive at the dual equation by taking advantage of both the BSM framework[7] and homogeneity of price functions. For this reason, it is called the Black-Scholes-Merton dual equation in Proposition 1. We believe the dual equation cannot be derived in any other way.[8] Our approach is unique, with two distinctive features. First, the strike is the variable in the dual equation. Second, the price function for options is, by definition, the same in both the dual and BSM equations.

It is tempting to believe that one can derive the BSM dual equation by changing the numeraire. This belief is incorrect for two reasons. First, changing numeraire, which deals only with the underlying dynamics (see, for example, Schroder, 1999), does not directly involve the strike price. Strike only appears in the terminal condition (i.e., the payoff function) or the solution to the BSM equation, however. Second, if it could be derived without considering the property of payoff, as is done in a change of numeraire, the dual equation would not subject to the functional constraint of homogeneity and thus be correct for all payoff functions. Remember, though, that Equation (1) is coupled to the BSM equation via price functions of homogeneous of degree one.

[7] It is easy to show that the dual equation does not work for arithmetic Brownian motions (ABM), because ABM is not scale-invariant.

[8] One might also argue from an anthropological perspective: if the dual equation is so obvious and trivial to derive, it must have appeared somewhere previously in publication or online. The lack of such evidence doubtlessly contradicts such naïve beliefs. The likely defense – it is so evident that nobody has ever bothered to work it out – sounds extremely weak, and might be viewed as the very reason why it has been overlooked so far.

Readers might erroneously think that the perceived "symmetry" in currency options invalidates our approach. It is easy to see that this is not the case. Denote the exchange rate by $F$ (i.e., the number of domestic currency versus one unit of foreign currency), and we have the well-known BSM equation:

$$\frac{\partial v}{\partial t} + (r - r_f)F\frac{\partial v}{\partial F} + \frac{1}{2}\sigma^2 F^2 \frac{\partial^2 v}{\partial F^2} = rv$$

where $r_f$ is the foreign risk-free interest rate. For the reciprocal $G = \frac{1}{F}$ (a.k.a., a change of numeraire), we can easily write out a second equation:

$$\frac{\partial u}{\partial t} + (r_f - r)G\frac{\partial u}{\partial G} + \frac{1}{2}\sigma^2 G^2 \frac{\partial^2 u}{\partial G^2} = r_f u$$

which also have the same form as the BSM equation, but does not involve the strike price. Further, the price functions $v$ and $u$ are unrelated and measured in different currencies. As we mentioned earlier, the dual and BSM equations share the same price function (and currency). Therefore, the second BSM equation for currency options is very different from the dual equation – it may sound pretentious, but the perspective via dual is entirely novel.

## 3. THE PUT-CALL EQUALITY

The dual nature of Equation (1) with respect to the BSM equation has intriguing implications for options' pricing. The results for a general put and its corresponding call in the world of the BSM dual equation are proved and summarized as follows.

**PROPOSITION 2:** Let's refer to the original BSM world as the BSM space. Further, the BSM dual equation world is named the dual space, in which the strike price is treated as the "underlying" variable and the current stock price is assumed to be the "strike", while the roles of *r* and $q$ are switched. Further, dual calls (puts) are used to mean call (put) options in the dual space. For options with payoffs that are homogenous of degree one (i.e., consistent with Proposition 1), denote the call prices by $V_C(S_0, r; K, q)$ and put by $V_P(S_0, r; K, q)$, in which the positions of the variables are significant and sequentially denote stock price, risk-free rate, strike price, and dividend yield:

(a) The price of a put $V_P(S_0, r; K, q)$ is equal to the price of its corresponding dual call $V_{\tilde{C}}(K, q; S_0, r)$. Note that we use ~ on top of the $C$ (or $P$) to indicate the dual space;

(b) The delta of the put $\frac{\partial V_P}{\partial S_0}$ is equal to $\frac{V_{\tilde{C}}}{S_0} - m\Delta$, where $\Delta = \frac{\partial V_{\tilde{C}}}{\partial K}$ is the delta of the corresponding dual call, $m = \frac{K}{S_0}$ is the moneyness. $S_0$ and *K* stand for stock price and strike price defined in the BSM space, respectively;

(c) The gamma of the put $\frac{\partial^2 V_P}{\partial S_0^2}$ is equal to $m^2\Gamma$, where $\Gamma = \frac{\partial^2 V_{\tilde{C}}}{\partial K^2}$ is the gamma of the corresponding dual call.

**PROOF**: (a) According to Proposition 1, $V_P(S_0, r; K, q)$, as a solution of the BSM equation, must solve Equation (1) too. Note that Equation (1) has the same form as the BSM equation. Further, if we treat *K* as stock price, *q* as risk-free rate, and *r* as dividend yield, Equation (1) and the BSM equation are precisely the same PDE. Therefore, under the dual equation, $V_P(S_0, r; K, q)$ is still an option price function of stock price *K*, risk-free rate *q*, and dividend *r*. Finally, in $V_P(S_0, r; K, q)$, *K* (i.e., the strike price) is the only parameter that does not appear in the BSM equation, while $S_0$ is the only parameter that is not in Equation (1). Obviously, $S_0$ under Equation (1) plays the role of strike price. To summarize, $V_P(S_0, r; K, q)$ under Equation (1) is the price of an option with *K* as stock price, *q* as risk-free rate, *r* as dividend yield, and $S_0$ as strike price.

Because *K* has switched the role with $S_0$, $V_P(S_0, r; K, q)$ is no longer the price of a put under the dual equation. Therefore, $V_P(S_0, r; K, q)$ now must be the price of a call, which we denote by $V_{\tilde{C}}(K, q; S_0, r)$.[9] To conclude, $V_P = V_{\tilde{C}}$.

(b) From (a), we have $\frac{\partial V_P}{\partial K} = \frac{\partial V_{\tilde{C}}}{\partial K}$. Combining with Equation (2), we obtain the result for the put delta.

[9] Our proof directly deals with the price function of options or a solution to the PDEs, and all boundary conditions are taken care of automatically, because, by definition, a solution is a result after considering all boundary conditions. Let's use the early exercise as an example. Assume that a put reaches its exercise boundary at a stock price of $S_t$, and we know that the boundary will be $V_P(S_t, r; K, q) = K - S_t$. Because $V_P(S_t, r; K, q) = V_{\tilde{C}}(K, q; S_t, r)$ is given, we have $V_{\tilde{C}}(K, q; S_t, r) = K - S_t$, which, with *K* as stock price and $S_t$ as strike price, is exactly the exercise boundary of the call in the dual space. Therefore, a one-to-one correspondence exists between the put and call exercise boundaries. We genuinely thank Steven Li for helping make the proof better.

(c) From (b), we have $\frac{\partial^2 V_P}{\partial K^2} = \frac{\partial^2 V_{\tilde{C}}}{\partial K^2}$. Utilizing Equation (3), we obtain the result for the put gamma. QED.

The interpretation of Proposition 2a is straightforward; trivially, any put can be priced as its corresponding call by simply swapping stock price (dividend yield) for strike price (risk-free rate) simultaneously. For example, a put with stock price 40, strike 50, risk-free rate 6%, and without dividends can be priced as a call with stock price 50, strike 40, risk-free rate 0%, and dividend yield 6%, or $V_P(40,6\%;50,0) = V_{\tilde{C}}(50,0;40,6\,\%)$ in mathematical notations. This will be easily seen later in Table 1 for American options.

The above proof is apparently correct also for any call and its corresponding dual put. With Proposition 2, it is clear that the price, delta, and gamma of a put (call) can be obtained from those of a call (put) in the dual space. This provides a new approach for understanding and pricing options, referred to as the put-call equality, and can be truly useful when the pricing in the BSM space seems challenging (see Sections 3.3 and 3.4 below). Notably, the proposition is general and works for a broad class of options with payoffs of homogeneous of degree one.

Furthermore, Proposition 2 or the put-call equality has appealing implications for implementing options' pricing algorithms. In principle, only one type of options, either calls or puts, is needed to be implemented in computer coding. If the pricing of call is implemented, puts can then be priced via the same codes by simply swapping stock price (dividend yield) for strike price (risk-free rate) simultaneously, and vice versa. This leads to simple and clean, small and efficient implementations of pricing algorithms in real-world applications.

It is worth noting that the practical applicability of Proposition 2 hinges on the analytic formulas for delta and gamma. With delta and gamma, we have a complete story for pricing and hedging. Undoubtedly, delta and gamma are two of the most important and useful Greeks. According to Chatterjea and Jarrow (2012), they are also the only valid Greeks. Without delta and gamma, a mere price relationship, such as Proposition 2a, would not be useful at best.

Just as the dual equation is novel, we believe the put-call equality (especially Proposition 2b-2c) is also new, because Proposition 2 directly follows from the dual equation. It is also tempting to think that the put-call equality is trivial and can be obtained via changing numeraire, at least for currency options. But this is not true. In fact, we cannot simply price one currency put as one currency call via changing numeraire, because a put with payoff, $\max(K - F_T, 0)$, is not equal to a call with payoff, $\max(G_T - X, 0) = \max(\frac{1}{F_T} - \frac{1}{K}, 0)$. Further, we have to manipulate $c\big(G_0, r_f; X, r\big) \times K$ first by working with the European call formula,[10] and then to make a currency conversion to obtain:

[10] "Show that the formula in equation (17.12) for a put option to sell one unit of currency A for currency B at strike price $K$ gives the same value as equation (17.11) for a call option to buy $K$ units of currency B for currency A at strike price $1/K$." (Problem 17.8, Page 380, Hull (2015)).

$$c\left(\frac{1}{F_0}, r_f; \frac{1}{K}, r\right) \times K \times F_0 = p\left(F_0, r; K, r_f\right)$$

This is hard, complicated, and logically backward.[11] By contrast, our put-call equality, as a direct result of the coupling between the dual and BSM equations and applicable to currency options as a special case, is as simple and clean as it can be, which no doubt stands out. Worse for changing numeraire, we cannot easily "see" whether the result is true for American or other options. Finally, it is unclear how to obtain analytic formulas for delta and gamma via changing numeraire. To the best of our knowledge, we believe that our "analytic" formulas (i.e., Proposition 2b-2c) are issues not addressed by previous researchers. Importantly, their derivation is far from trivial without utilizing the dual equation.

For illustration, the put-call equality is verified in the following for European options via the Black-Scholes-Merton formulas and numerically for American options. Further, the put-call equality is utilized to shed light on the complex issues of pricing certain American puts and known dollar dividends in options' pricing.

### 3.1. European Options

Denote the European call (put) price by $c$ ($p$). The well-known Black-Scholes-Merton put formula is:

$$p(S_0, r; K, q) = Ke^{-rT}N(-d_2) - S_0e^{-qT}N(-d_1)$$

[11] Note that the proof of "put-call symmetry" for European options in Detemple (1999) is also "backward" for relying on the BSM formulas.

$$d_1 = \frac{\ln(\ S_0/K) + (r - q + 0.5\sigma^2)T}{\sigma\sqrt{T}}, d_2 = d_1 - \sigma\sqrt{T}$$

Further, the corresponding dual call price is as follows, with $S_0$ ($r$) switching places with $K$ ($q$) within the original BSM call formula:

$$\tilde{c}(K, q; S_0, r) = Ke^{-rT}N(d_1^*) - S_0 e^{-qT}N(d_2^*)$$

$$d_1^* = \frac{\ln(\ K/S_0) + (q - r + 0.5\sigma^2)T}{\sigma\sqrt{T}}, d_2^* = d_1^* - \sigma\sqrt{T}$$

It is easy to verify that $\tilde{c}(K, q; S_0, r)$ is a solution of Equation (1).

Trivially $d_1^* = -d_2$ and $d_2^* = -d_1$. Consequently, $\tilde{c} = p$. Similarly, the corresponding dual put price can be shown to be the same as the call price. Therefore, Proposition 2a is verified for European options. It is worth pointing out that the proof of $\tilde{c}(K, q; S_0, r) = p(S_0, r; K, q)$ in Detemple (1999) relies on the BSM formulas and therefore is logically "backward" by contrast.

The put delta is known to be $-e^{-qT}N(-d_1)$. Further, it is not hard to see that the call delta in the dual space is $\Delta = e^{-rT}N(d_1^*) = e^{-rT}N(-d_2)$. The first term of $\frac{V_{\tilde{c}}}{S_0}$ turns out to be the same as $m\Delta$ and cancels out, while the second term of $\frac{V_{\tilde{c}}}{S_0}$ is $-e^{-qT}N(d_2^*) = -e^{-qT}N(-d_1)$, which is exactly the put delta the BSM space. Therefore, Proposition 2b is also verified for European options. Similarly, Proposition 2c can be verified for European options as well.

Of course, the European call and put are related via the well-known put-call parity (Black and Scholes, 1973), so the put-call equality may seem to be unnecessary for European options. Nevertheless, the put-call equality arguably is significantly more general than the put-call parity.

In addition to European options, the put-call equality also works for American and many other options.

### 3.2. American Equity Options

For American options when early-exercise can be optimal, the put-call equality can only be verified via numerical examples. The numerical parameters utilized in this section are assumed as follows. The risk-free rate is 6%, the volatility is 40%, and the maturity of options is 365 days. The numerical pricing is carried out via binomial tree with daily steps here. Note that even though the chosen numerical values in this section are quite limited, the conclusions are nevertheless valid in general.

It is well-known that when the underlying stock does not pay any dividend, an American equity call should never be exercised early (Merton, 1973), and thus can be priced by the Black-Scholes-Merton call formula. Further, the premium of early-exercising can usually be ignored approximately (see a heuristic argument in Appendix 2). On the other hand, it may be optimal to exercise an American put before maturity. Therefore, it is only necessary to verify Proposition 2 for American puts. Table 1 shows clearly that American puts can be priced as American calls in the dual space with negligible errors.

Table 1 here

In addition, the delta and gamma of an American put option can be computed accurately from the delta and gamma of its corresponding dual call, respectively. Most of the corresponding deltas match to four digits after the decimal point, and so do gammas (Table 2).

Table 2 here

### 3.3. American Currency Puts

Ever since Merton (1973), it has been believed that American puts can only be priced numerically. The put-call equality leads to new insights on American currency puts, however. If the domestic risk-free rate were zero, the corresponding dual call of the American put would be equivalent to an American equity call without dividends, and thus not be exercised early at all. Applying the dual pricing reasoning again but in reverse, we know that this dual call has the same price as its corresponding European put in the BSM space. Therefore, the American currency put with a zero domestic risk-free rate can then be priced exactly by the BSM put formula.[12] This is certainly more accurate than numerical approaches, all of which can unfortunately be prone to numerical errors.

One might argue that risk-free interest rates are rarely zero. Still, it is not uncommon for $r < r_f$. In this case, the early-exercising premium of an American put can be negligible, as the put-call equality reveals once more. In other words, American puts with $r < r_f$ can be priced approximately by the BSM put formula.

[12] The "BSM formulas" here loosely refer to pricing formulas for European currency options in Garman and Kohlhagen (1983).

The above analyses for American puts are easily confirmed numerically (Table 3). When the domestic risk-free rate is zero, the prices of American puts and their corresponding European puts are almost the same, provided that numerical errors exist in the binomial-tree pricing of American puts. With a domestic risk-free rate of 3%, which is one half of the foreign risk-free rate, the absolute pricing errors of the BSM put formula for the American puts are all below 0.1% and still truly small. Finally with a domestic risk-free rate as high as 5%, the biggest absolute error is nevertheless under 1% and perhaps well acceptable in practice.

Table 3 here

### 3.4. Known Dollar Dividends

The issue of known dollar dividends in options' pricing is complicated and seems to be still an open question. In the popular textbook of Hull (2015), European options are priced by applying the BSM formulas to the risky component of the stock process. With the notation used in Section 3.1, the price of a European call is then $c(S_0 - D, r; K, 0)$, where $D$ is the present value of dollar dividends, and the parameter 0 indicates a zero dividend yield or without dividends. Obviously, the risky component of the stock process is modeled as GBM (and thus scale-invariant), and the BSM formula $c(S_0 - D, r; K, 0)$ is homogenous of degree one with respect to $S_0 - D$and $K$, and is a solution of the dual equation. Note that Hull (2015) states explicitly that this approach has been criticized by some recent literature (for example, Areal and Rodrigues, 2013; Guo and

Liu, 2019).

Remarkably, whether the Hull (2015) approach is legitimate can be analyzed via the put-call equality. If we follow the Hull approach for American calls, the American price is then $C(S_0 - D, r; K, 0)$. Even though it is perhaps unclear how to handle early exercise for calls here, the corresponding put in the dual space turns out to be easier to analyze. The price of the dual American put is simply $\tilde{P}(K, 0; S_0 - D, r)$, where the underlying variable $K$ represents again the risky component. Since $S_0 - D$ is now the strike (and a constant), $D$ is absent from the underlying process of the risky component, and the "risk-free rate" is zero, the dual put will not be exercised early (see Section 3.3 above). Therefore, $\tilde{P}$ is equal to the price of the dual European put. Because the price of the dual European put is the same as $c(S_0 - D, r; K, 0)$, we have $C(S_0 - D, r; K, 0) = c(S_0 - D, r; K, 0)$. In other words, the American call would have the same price as the corresponding European call under the Hull (2015) approach. On the other hand, it is known that it can be optimal to exercise the call right before the dividend payment (Merton, 1973). Therefore, we have a contradiction here. This implies that the assumption of the risky component is problematic. To conclude, the textbook treatment of dollar dividends for European options is probably problematic.

A moment of reflection confirms this unfortunate conclusion. Indeed, if one works only with the risky component of the stock process, there is no place for any known dividend; the present value of the dividend can only be deducted from the initial stock price. On the other hand, the risky component of the stock process should be independent of and indifferent to the type of the option, be it European or American. In other words, the risky component of the stock process should work in the same way for both European and American options. Without any mechanism for early-exercising, an American call thus has to be the same as its corresponding European call.

In summary, the implication of the put-call equality on options' pricing is very innovative and insightful. Further, the put-call equality leads to an efficient evaluation via the BSM put formula for American currency puts, when the foreign risk-free rate is higher. Finally, a new understanding can be gained on Hull's textbook approach to known dividends in pricing European options, as the above dual pricing analyses reveal.

*2.5. Versus the Put-call Symmetry*

Detemple (1999) provides a painstaking and detailed case-by-case treatment of the put-call symmetry for various types of options under geometric Brownian motions and more general underlying processes. It is reassuring that our Proposition 2a confirms the primary (and no doubt the most commonly used) parts of his extensive analyses, even though our approach is radically different.

Arguably, our "new" result in Propositions 2a can be worthwhile in itself from a scientific perspective. Unlike Detemple (1999), we utilize a totally different approach to show trivially via the BSM dual equation the equality of puts and their dual calls for a broad class of options with payoffs of homogeneous of degree one. Obviously, our unique proof is significantly simpler, considerably shorter, and much easier to understand. Further, our alternative approach can lead to new understandings of and insights into existing issues, as were shown in Sections 3.3 and 3.4.

More importantly, the analytic formulas for delta and gamma finally elevate the idea of the put-call price relationships to true practical applicability, as were shown in Sections 3.1 and 3.2 for example for European and American options. Proposition 2 shows unambiguously that even though the price of a put (call) in the BSM space is the same as that of the call (put) in the dual space, the Greeks delta and gamma are not. Consequently, delta and gamma cannot be computed directly in the dual space. Detemple (1999) addresses only the issue of pricing, however. Without providing solutions to delta and gamma, the two most important hedge parameters, Detemple's put-call symmetry appears to be lame and not truly useful.

## 4. FUTURES OPTIONS

Futures options are very popular worldwide. It is known that futures can be regarded as underlying assets that pay the risk-free rate as continuous dividend yields (Hull, 2015). When the risk-free rate is equal to the continuous dividend yield, the dual space becomes very simple and the dual reasoning based on Proposition 2 can lead to insights on the pricing of futures options. Actually, we have the following proposition.

**PROPOSITION 3:** Assume the risk-free rate and continuous dividend yield are the same (and thus interchangeable), so that we can drop the placeholders for *r* and $q$ in mathematical notations here. Define $m = \frac{K}{S_0}$. For options with payoffs that are homogenous of degree one (i.e., consistence with Proposition 2):

(a) The price of a put $V_P(S_0; K)$, with futures price $S_0$ and strike $K$, is equal to $\frac{1}{m} V_C(m^2 S_0 ; K)$;

(b) The delta of the put is equal to $\frac{V_C}{K} - m\Delta_C$, where $\Delta_C$ is the delta of the call;

(c) The gamma of the put is equal to $m^3 \Gamma_C$, where $\Gamma_C$ is the gamma of the call.

**PROOF**: (a) Since the price is homogenous of degree one for an option with a payoff of homogenous of degree one, we have for a put:

$$mV_P(S_0; K) = V_P(K; m^2 S_0)$$

Because *r* equals $q$, the right-hand side can be regarded as the put price in the dual space. Therefore, we can apply Proposition 2a to obtain:

$$V_P(K; m^2 S_0) = V_C(m^2 S_0; K)$$

Combining the two equations above, we have $V_C(m^2 S_0; K) = mV_P(S_0; K)$ and Proposition 3a is proved. Note that trivially, we will obtain $V_P(m^2 S_0; K) = mV_C(S_0; K)$ if we start with a call and reverse the proof process.

(b) Writing *m* out explicitly and applying the chain rule, we have the put delta from (a) as

follows:[13]

$$\frac{\partial V_P}{\partial S_0} = \frac{1}{\partial S_0}\left[\frac{S_0}{K} V_C(K^2 S_0^{-1})\right]$$
$$= \frac{1}{K}\left[V_C - S_0 \frac{\partial V_C}{\partial (K^2 S_0^{-1})} K^2 S_0^{-2}\right]$$
$$= \frac{V_C}{K} - m\Delta_C$$

(c) From (b), the gamma of the put is then:

$$\frac{\partial^2 V_P}{\partial S_0^2} = \frac{1}{\partial S_0}[K^{-1} V_C(K^2 S_0^{-1}) - K S_0^{-1} \Delta_C(K^2 S_0^{-1})] = m^3 \Gamma_C$$

QED.

Proposition 3 shows that given *q* equals *r*, the price, delta, and gamma of a put (call) can be obtained simply from those of a corresponding call (put) for a broad class of options with payoffs of homogeneous of degree one. One might call this a "generalized" put-call parity.

While Proposition 2 relates put prices in the BSM space to call prices in the dual space, Proposition 3 deals with put and call prices in the same BSM space. As was argued previously, Proposition 2 means that only calls have to be implemented by coding. Proposition 3 implies that for any call priced, a corresponding put can be priced for free. Trivially, we can see this from $V_P(m^2 S_0; K) = m V_C(S_0; K)$; the price of put at an underlying price $m^2 S_0$ is *m* times the price of call at the underlying price $S_0$. The put delta and gamma can be obtained similarly via the proof of Proposition 3b and 3c.

[13] For notational brevity, we drop the strike from the price functions of options in the proof.

Further, we have the following simple price relationships.

**Corollary 3.1** With the same assumptions and notations as in Proposition 3:

(a) The price of a put $V_P(S_0; K)$ is equal to $V_C(K; S_0)$;

(b) The delta of the put is equal to $\frac{V_C}{S_0} - m\Delta_C$, where $\Delta_C = \frac{\partial V_C}{\partial K}$;

(c) The gamma of the put is equal to $m^2\Gamma_C$, where $\Gamma_C = \frac{\partial^2 V_C}{\partial K^2}$.

**PROOF**: (a) From Proposition 3a and the option price of homogenous of degree one, we have:

$$V_P(S_0; K) = \frac{1}{m} V_C(m^2 S_0; K) = V_C(K; S_0)$$

Note that also trivially $V_C(S_0; K) = V_P(K; S_0)$.

(b) From (a), we have $\frac{\partial V_P}{\partial S_0} = \frac{\partial V_C}{\partial S_0}$. Combining with Equation (2), we obtain the put delta.

(c) From (b), we have $\frac{\partial^2 V_P}{\partial S_0^2} = \frac{\partial^2 V_C}{\partial S_0^2}$. Combining with Equation (3), we obtain the put gamma.

QED.

Corollary 3.1 shows that the prices of futures call and put are "symmetric" with respect to the underlying price and strike price. Similar to Proposition 3, the prices of calls and puts in the BSM space are related for futures options; further, we obtain a put price for free, whenever we price a call option.

Finally, we have the price relationships between futures calls in Corollary 3.2.

**Corollary 3.2** Let $C'$ denotes the call with a futures price $K$ and a strike price $S_0$ and define $m = \frac{K}{S_0}$. From Proposition 3 and Corollary 3.1:

(a) $V_{C'}(K; S_0) = \frac{1}{m} V_C(m^2 S_0; K)$;

(b) $\Delta_{C'} = \Delta_C$;

(c) $\Gamma_{C'} = m \ \Gamma_C$.

The price relationships between futures puts can be obtained in exactly the same way. Again, we can verify Proposition 3 in the following for European options via the Black-Scholes-Merton formulas and numerically for American options, as examples. The more complex options are left out for brevity.

**4.1. European Futures Options**

Proposition 3a can be trivially verified and Proposition 3c is similar to delta, so we will only work out the detail for Proposition 3b here. The delta of put with underlying price $S_0$ and strike $K$ is known as $-e^{-qT}N(-d_1)$, as was mentioned previous in Section 3.1. With underlying price $m^2 S_0$ and strike $K$, we can easily show that $\Delta_C = e^{-qT}N(d_1') = e^{-qT}N(-d_2)$, because $d_1' = -d_2$, and $d_2' = -d_1$. Now it is not hard to see that the first term of $\frac{V_C}{K}$ cancels out $-m\Delta_C$, with the remaining being the second term of $\frac{V_C}{K}$, or $-e^{-qT}N(-d_1)$. Thus, Proposition 3b is verified for European options.

**4.2. American Futures Options**

Again, we use binomial tree to verify Proposition 3 and Corollary 3.1 numerically for American futures options. The numerical parameters utilized here are the same as in Section 3. The risk-free rate is 6%, the volatility is 40%, and the maturity of options is 365 days. The numerical pricing is carried out via binomial tree with daily steps.

Table 4 shows prices, deltas, and gammas for American futures options via Proposition 3. Undoubtedly, American puts can be priced as American calls, with all the reported values match to four digits after the decimal point.

Table 4 here

Prices, deltas, and gammas for American futures options via Corollary 3.1 are summarized in Table 5. All the errors are tiny, so that prices, deltas, and gammas are all numerically the same for practical purpose.

Table 5 here

Actually, we can verify Corollary 3.2 from Tables 4 and 5. For example, $V_{C'}(40;36) = 7.881$ in Table 5 and $V_C(44.444;40) = 8.757$ in Table 4, respectively. Note that here $m = 1.1111$. The two corresponding deltas are both 0.6562. Lastly, $\Gamma_{C'} = 0.0226$ and $\Gamma_C = 0.0204$. The rest of other four cases in Tables 4 and 5 can all be verified as well.

To summarize, Proposition 3 and its two corollaries imply that with one futures call priced, we can obtain the price of a put and of a second call for free. Similarly, it can be shown that with one futures put priced, we can obtain the price of a call and of a second put for free. These price properties between calls and puts or among calls (puts) are novel and interesting.

## 5. CONCLUSIONS

We derive the Black-Scholes-Merton dual equation, a second-order partial differential equation with respect to the option's strike price. This new equation is very general and works for European, American, Bermudan, Asian, barrier, lookback, etc. options.

The BSM dual equation shows important implications for options' pricing and hedging. With the dual equation, it is easily proved that the price of put (call) options are equal to those of the corresponding call (put) options in the dual space; further, the deltas of these options are related via a simple analytic formula, so are gammas. These relations are readily verified for European options via the BSM formulas and for American options via binomial-trees. The dual nature implies that American currency puts can be valued efficiently by the BSM formula of the dual call, when the foreign risk-free rate is higher. Further, the dual argument seems to suggest that the treatment of known dollar dividends in European options' pricing is problematic in the popular textbook of Hull (2015).

The practical implication of the put-call equality is really appealing. Under the put-call equality, only call options have to be programmed in coding, because the price, delta, and gamma of puts can all be computed via the corresponding dual calls. The resultant algorithms for pricing and hedging will be simple, clean, and more efficient. For futures options, the put-call equality results in novel price properties between calls and puts or among calls (puts).

Future researches could proceed at least in two directions. First, the dual pricing reasoning may lead to new insights into the early-exercising boundary for American options. Second, it can be interesting to extend the dual equation to stochastic volatility and/or stochastic risk-free rate models.

## APPENDIX

### 1. Derivation of Equation 3

It is known that the first-order partial derivatives of functions of homogeneous of degree one, such as $v(S_t,K)$, are homogeneous of degree zero:

$$\frac{\partial v(\beta S_t, \beta K)}{\partial(\beta S_t)} = \frac{\partial v(S_t, K)}{\partial S_t}, \frac{\partial v(\beta S_t, \beta K)}{\partial(\beta K)} = \frac{\partial v(S_t, K)}{\partial K}$$

where $\beta$ is an arbitrary positive real variable. Differentiating both sides of these two equations with respect to $\beta$, setting $\beta$ to 1, and eliminating the mixed second-order partial derivative, we obtain Equation (3) as follows:

$$S_t^2 \frac{\partial^2 v}{\partial S_t^2} = K^2 \frac{\partial^2 v}{\partial K^2}$$

### 2. Early Exercise of American Calls

In the binomial tree model, the risk-neutral probability of the underlying price going up is (Hull, 2015):

$$Q = \frac{\exp(\ [r-q]\delta) - d}{u - d}$$

where $\delta$ is the time interval, and $u$ ($d$) is the up (down) parameter. The hold value of American call is at least:[14]

$$e^{-r\delta}[Q(Su-K) + (1-Q)(Sd-K)] = Se^{-q\delta} - Ke^{-r\delta}$$

$$\approx (S-K) - (Sq-Kr)\delta$$

[14] The hold value will accumulate and increase going backwards in time.

Apparently, it is optimal to hold the call if the dividend yield is zero. Further, it is approximately optimal to hold the call with $Sq < Kr$. Since $q < r$ usually, only deep in-the-money American calls will be exercised early. Therefore, the early-exercising premium can be ignored approximately. On the other hand, the call shall be exercised early if $Sq > Kr$.

## REFERENCES

Alexander, C., & Nogueira, L. (2006). Hedging options with scale-invariant models, *Working Paper*, University of Reading.

Andreasen, J. (1997). Essays in contingent claims pricing, *Ph.D. Thesis*, University of Aarhus.

Andreasen, J., & Carr, P. (2002) Put call reversal, *Working Paper*, Bank of America.

Areal, N., & Rodrigues, A. (2013) Fast trees for options with discrete dividends, *Journal of Derivatives* 21, 49-63.

Bates, D. (1988). The crash premium: Option pricing under asymmetric processes, with applications to options on Deutschemark futures, *Working Paper*, University of Pennsylvania.

Black, F., & Scholes, M. (1973). The pricing of options and corporate liabilities, *Journal of Political Economy* 81, 637-654.

Breeden, D. T., & Litzenberger, R. H. (1978) Prices of state-contingent claims implicit in option prices, *Journal of Business* 51, 621-651.

Carr, P., & Hirsa, A. (2003). Why be backward? Forward equations for American options, *Risk* 9, 103-107.

Chatterjea, A., and Jarrow, R. A. (2012). The Dangers of calibration and hedging the Greeks in option pricing, *Journal of Financial Education* 38, 1-12.

Detemple, J. (1999.) American options: Symmetry properties, *Working Paper*, Boston University and CIRANO.

Dupire, B. (1993). Pricing and hedging with smiles, *Proceedings of AFFI Conference* La Boule, June 1993.

Garman, M. B., & Kohlhagen, S. W. (1983). Foreign currency option values, *Journal of International Money and Finance* 2, 231-237.

Gatheral, J. (2002). Lecture 1: Stochastic volatility and local volatility. *Case Studies in Financial Modelling Course Notes*, Courant Institute of Mathematical Sciences.

Grabbe, J. O. (1983). The pricing of call and put options on foreign exchange, *Journal of International Money and Finance* 2, 239-253.

Guo, S., & Liu, Q. (2019). A simple accurate binomial tree for pricing options on stocks with known dollar dividends, *Journal of Derivatives* 26, 54-70.

Guo, S., & Liu, Q. (2023). Improving and extending the Wu-Zhu static hedge, *Journal of Derivatives* 26, Spring (forthcoming).

Haug, E. (2002). A look in the antimatter mirror, *www.wilmott.com*, January.

Hull, J. C. (2015). *Options, Futures, and Other Derivatives*, 9th ed. (Prentice Hall, Upper Saddle River, NJ).

Merton, R. C. (1973). Theory of rational option pricing, *Bell Journal of Economics and*

*Management Science* 4, 141-183.

Peskir G., & Shiryaev, A. (2002). A note on the call-put parity and a call-put duality, *Theory of Probability & its Applications 46*, 167-170.

Schroder, M. (1999). Changes of numeraire for pricing futures, forwards, and options, *Review of Financial Studies* 12, 1143-1163.

Wu, L., & Zhu, J. (2016). Simple robust hedging with nearby contracts, *Journal of Financial Econometrics* 15, 1-35.

**Table 1** Binomial tree pricing of American options.

| American Put | | | American Call | | | |
|---|---|---|---|---|---|---|
| $r = 0.06, q = 0$ | | | $r = 0, q = 0.06$ | | | |
| S | K | P | S | K | C | Error (%) |
| 36 | 40 | 7.108 | 40 | 36 | 7.109 | 0.015 |
| 38 | 40 | 6.157 | 40 | 38 | 6.158 | 0.019 |
| 40 | 40 | 5.322 | 40 | 40 | 5.323 | 0.022 |
| 42 | 40 | 4.592 | 40 | 42 | 4.593 | 0.025 |
| 44 | 40 | 3.955 | 40 | 44 | 3.956 | 0.029 |

*Notes*: S: stock price. K: strike price. P: put price. C: call price. Error: $C/P - 1$. The annual volatility is 40%. The options mature in 365 days. The binomial tree has daily steps. $r$ is the annual risk-free rate and $q$ is the annual dividend yield.

**Table 2** Delta and gamma of American puts.

| American Put | | | American Call | | | | | |
|---|---|---|---|---|---|---|---|---|
| $r = 0.06, q = 0$ | | | $r = 0, q = 0.06$ | | | | | |
| S | K | delP | S | K | C | delC | Delta | Error (%) |
| 36 | 40 | -0.5089 | 40 | 36 | 7.109 | 0.6357 | -0.5088 | -0.006 |
| 38 | 40 | -0.4467 | 40 | 38 | 6.158 | 0.5783 | -0.4467 | -0.003 |
| 40 | 40 | -0.3908 | 40 | 40 | 5.323 | 0.5239 | -0.3908 | -0.001 |
| 42 | 40 | -0.3407 | 40 | 42 | 4.593 | 0.4726 | -0.3407 | 0.002 |
| 44 | 40 | -0.2962 | 40 | 44 | 3.956 | 0.4247 | -0.2962 | 0.005 |
| S | K | gamP | S | K | | gamC | Gamma | Error (%) |
| 36 | 40 | 0.0326 | 40 | 36 | | 0.0264 | 0.0325 | -0.024 |
| 38 | 40 | 0.0295 | 40 | 38 | | 0.0266 | 0.0295 | -0.023 |
| 40 | 40 | 0.0265 | 40 | 40 | | 0.0265 | 0.0265 | -0.021 |
| 42 | 40 | 0.0236 | 40 | 42 | | 0.0260 | 0.0236 | -0.019 |
| 44 | 40 | 0.0210 | 40 | 44 | | 0.0254 | 0.0210 | -0.017 |

*Notes*: S: stock price. K: strike price. C: call price. delP: put delta from binomial tree. delC: call delta from binomial tree. Delta: for put via Proposition 2b. gamP: put gamma from binomial tree. gamC: call gamma from binomial tree. Gamma: for put via Proposition 2c. Error: $\text{Delta}/\text{delP} - 1$ or $Gamma/gamP - 1$. The annual volatility is 40%. Both options mature in 365 days. The binomial tree has daily steps. $r$ is the annual risk-free rate and $q$ is the annual dividend yield.

**Table 3** Approximating American currency puts as European calls.

| American Put | | | European Call | | | |
|---|---|---|---|---|---|---|
| $r=0, r_f=0.06$ | | | $r=0.06, r_f=0$ | | | |
| F | K | P | F | K | c | Error (%) |
| 36 | 40 | 9.389 | 40 | 36 | 9.391 | 0.029 |
| 38 | 40 | 8.344 | 40 | 38 | 8.341 | -0.035 |
| 40 | 40 | 7.394 | 40 | 40 | 7.389 | -0.064 |
| 42 | 40 | 6.535 | 40 | 42 | 6.531 | -0.060 |
| 44 | 40 | 5.763 | 40 | 44 | 5.761 | -0.038 |
| $r=0.03, r_f=0.06$ | | | $r=0.06, r_f=0.03$ | | | |
| F | K | P | F | K | c | Error (%) |
| 36 | 40 | 8.545 | 40 | 36 | 8.543 | -0.025 |
| 38 | 40 | 7.555 | 40 | 38 | 7.549 | -0.083 |
| 40 | 40 | 6.660 | 40 | 40 | 6.653 | -0.105 |
| 42 | 40 | 5.856 | 40 | 42 | 5.850 | -0.093 |
| 44 | 40 | 5.138 | 40 | 44 | 5.135 | -0.063 |
| $r=0.05, r_f=0.06$ | | | $r=0.06, r_f=0.05$ | | | |
| F | K | P | F | K | c | Error (%) |
| 36 | 40 | 8.081 | 40 | 36 | 8.008 | -0.902 |
| 38 | 40 | 7.113 | 40 | 38 | 7.051 | -0.870 |
| 40 | 40 | 6.244 | 40 | 40 | 6.193 | -0.814 |
| 42 | 40 | 5.467 | 40 | 42 | 5.427 | -0.735 |
| 44 | 40 | 4.777 | 40 | 44 | 4.747 | -0.646 |

*Notes*: F: number of domestic currency per one unit of foreign currency. K: strike price. P: American put price. c: European call price. Error: $c/P-1$. The American put is priced via binomial tree with daily steps, while the European call is priced via the Black-Scholes-Merton formula. The annual volatility is 40%. Both options mature in 365 days. $r$ $(r_f)$ is the annual domestic (foreign) risk-free rate.

**Table 4** American futures puts via Proposition 3.

| American Put | | | American Call | | | | | |
|---|---|---|---|---|---|---|---|---|
| F | K | P | F | K | C | | Price | Error (%) |
| 36 | 40 | 7.881 | 44.444 | 40 | 8.757 | | 7.881 | 0.002 |
| 38 | 40 | 6.920 | 42.105 | 40 | 7.285 | | 6.920 | 0.003 |
| 40 | 40 | 6.061 | 40.000 | 40 | 6.061 | | 6.061 | 0.003 |
| 42 | 40 | 5.295 | 38.095 | 40 | 5.043 | | 5.295 | 0.003 |
| 44 | 40 | 4.618 | 36.364 | 40 | 4.198 | | 4.618 | 0.003 |
| F | K | delP | F | K | C | delC | Delta | Error (%) |
| 36 | 40 | -0.5102 | 44.444 | 40 | 8.757 | 0.6562 | -0.5102 | 0.000 |
| 38 | 40 | -0.4560 | 42.105 | 40 | 7.285 | 0.6062 | -0.4560 | 0.001 |
| 40 | 40 | -0.4058 | 40.000 | 40 | 6.061 | 0.5573 | -0.4058 | 0.001 |
| 42 | 40 | -0.3596 | 38.095 | 40 | 5.043 | 0.5100 | -0.3596 | 0.001 |
| 44 | 40 | -0.3175 | 36.364 | 40 | 4.198 | 0.4647 | -0.3175 | 0.002 |
| F | K | gamP | F | K | | gamC | Gamma | Error (%) |
| 36 | 40 | 0.0279 | 44.444 | 40 | | 0.0204 | 0.0279 | -0.003 |
| 38 | 40 | 0.0261 | 42.105 | 40 | | 0.0224 | 0.0261 | -0.002 |
| 40 | 40 | 0.0241 | 40.000 | 40 | | 0.0241 | 0.0241 | -0.002 |
| 42 | 40 | 0.0221 | 38.095 | 40 | | 0.0256 | 0.0221 | -0.001 |
| 44 | 40 | 0.0200 | 36.364 | 40 | | 0.0267 | 0.0200 | -0.001 |

*Notes*: F: futures price. K: strike price. P: put price. C: call price. Price: for put via Proposition 3a. delP: put delta. delC: call delta. Delta: for put via Proposition 3b. gamP: put gamma. gamC: call gamma. Gamma: for put via Proposition 3c. Error: $Price/P-1$, $\mathrm{Delta}/\mathrm{delP}-1$ or $Gamma/gamP-1$. The annual volatility is 40% and the annual risk-free rate is 6%. Both options mature in 365 days. The binomial tree has daily steps.

**Table 5** American futures puts via Corollary 3.1.

| American Put | | | American Call | | | | | |
|---|---|---|---|---|---|---|---|---|
| F | K | P | F | K | C | | | Error (%) |
| 36 | 40 | 7.881 | 40 | 36 | 7.881 | | | 0.000 |
| 38 | 40 | 6.920 | 40 | 38 | 6.920 | | | 0.000 |
| 40 | 40 | 6.061 | 40 | 40 | 6.061 | | | 0.000 |
| 42 | 40 | 5.295 | 40 | 42 | 5.295 | | | 0.000 |
| 44 | 40 | 4.618 | 40 | 44 | 4.618 | | | 0.000 |
| F | K | delP | F | K | C | delC | Delta | Error (%) |
| 36 | 40 | -0.5102 | 40 | 36 | 7.881 | 0.6562 | -0.5102 | 0.000 |
| 38 | 40 | -0.4560 | 40 | 38 | 6.920 | 0.6062 | -0.4560 | 0.000 |
| 40 | 40 | -0.4058 | 40 | 40 | 6.061 | 0.5573 | -0.4058 | 0.000 |
| 42 | 40 | -0.3596 | 40 | 42 | 5.295 | 0.5100 | -0.3596 | 0.000 |
| 44 | 40 | -0.3175 | 40 | 44 | 4.618 | 0.4647 | -0.3175 | 0.000 |
| F | K | gamP | F | K | | gamC | Gamma | Error (%) |
| 36 | 40 | 0.0279 | 40 | 36 | | 0.0226 | 0.0279 | 0.000 |
| 38 | 40 | 0.0261 | 40 | 38 | | 0.0235 | 0.0261 | 0.000 |
| 40 | 40 | 0.0241 | 40 | 40 | | 0.0241 | 0.0241 | 0.000 |
| 42 | 40 | 0.0221 | 40 | 42 | | 0.0243 | 0.0221 | 0.000 |
| 44 | 40 | 0.0200 | 40 | 44 | | 0.0243 | 0.0200 | 0.000 |

*Notes*: F: futures price. K: strike price. P: put price. C: call price. delP: put delta. delC: call delta. Delta: for put via Corollary 3.1b. gamP: put gamma. gamC: call gamma. Gamma: for put via Corollary 3.1c. Error: $C/P-1$, $\mathrm{Delta}/\mathrm{delP}-1$ or $Gamma/gamP-1$. The annual volatility is 40% and the annual risk-free rate is 6%. Both options mature in 365 days. The binomial tree has daily steps.